\documentclass{iaus}
\usepackage{graphics}

  \checkfont{eurm10}
  \iffontfound
    \IfFileExists{upmath.sty}
      {\typeout{^^JFound AMS Euler Roman fonts on the system,
                   using the 'upmath' package.^^J}%
       \usepackage{upmath}}
      {\typeout{^^JFound AMS Euler Roman fonts on the system, but you
                   dont seem to have the}%
       \typeout{'upmath' package installed. iaus.cls can take advantage
                 of these fonts,^^Jif you use 'upmath' package.^^J}%
      }
  \else
  \fi

  \checkfont{msam10}
  \iffontfound
    \IfFileExists{amssymb.sty}
      {\typeout{^^JFound AMS Symbol fonts on the system, using the
                'amssymb' package.^^J}%
       \usepackage{amssymb}%
         \let\leq=\leqslant
         
      }{}
  \fi

  \IfFileExists{amsbsy.sty}
    {\typeout{^^JFound the 'amsbsy' package on the system, using it.^^J}%
     \usepackage{amsbsy}}
    {\providecommand\boldsymbol[1]{\mbox{\boldmath $##1$}}}

\newsavebox{\astrutbox}
\sbox{\astrutbox}{\rule[-5pt]{0pt}{20pt}}

\title[Outskirts of Galaxy Clusters: intense life in the suburbs]
      {Galaxies in Present-day Clusters: 
Evolutionary Constraints from Their Distributions and Kinematics}

\author[A. Biviano \& P. Katgert]%
{Andrea Biviano$^1$%
\and Peter Katgert$^2$}

\affiliation{$^1$INAF/Osservatorio Astronomico di Trieste, 
via G. B. Tiepolo 11, I-34131 Trieste, Italy\\
$^2$Sterrewacht Leiden, Postbus 9513, Niels Bohrweg 2, 2300 RA Leiden,
The Netherlands}

\pubyear{2004}
\volume{195}
\pagerange{1--8}
\date{?? and in revised form ??}
\jname{Outskirts of Galaxy Clusters: intense life in the suburbs}
\editors{A. Diaferio, ed.}
\begin{document}

\maketitle

\begin{abstract}
We discuss evidence in local, present-day clusters of galaxies (from
the ENACS survey) about the way in which those clusters have evolved
and about the evolutionary relationships between the galaxies of
different morphological types in them. This evidence is complementary
to that obtained from the study of clusters at intermediate and high
redshifts. We argue that the spatial distribution and the kinematics
of the various types of galaxies in and outside substructures support
the following picture.

The {\em elliptical and S0 galaxies} have been around for a long time
and have obtained an isotropic velocity distribution. The spatial
distribution and kinematics of the {\em early spirals} are consistent
with the idea that many of their kind have transformed into an S0, but
that they have survived, most likely because of their velocities. The
distribution and kinematics of the {\em late spirals} are consistent
with a picture in which they have been accreted fairly recently. They
have mildly radial orbits and hardly populate the central regions,
most likely because they suffer tidal disruption. Finally, the
distribution and kinematics of the {\em galaxies in substructures,} when
taken at face value, imply tangential velocity anisotropy for these
galaxies, but this result may be (partly) due to the procedure by
which these galaxies are selected. A first attempt to take the effects
of selection into account shows that isotropic (or even mildly radial)
orbits of subcluster galaxies cannot be excluded.
\end{abstract}

\firstsection 
\section{Introduction and summary of the data sample}
\label{s-intro}
By observing the precursors of present-day clusters, it has become
possible to study in a direct manner the evolution of the clusters
themselves, and the evolutionary relationships between the various
types of galaxies in them. Images of rich clusters at redshifts $z
\approx 1$ show a high merger fraction, which is taken to be direct
evidence for the formation of elliptical galaxies by merging. Yet, it
is unlikely that {\em all} early-type galaxies formed around or before
that epoch (\cite[van Dokkum \& Franx 2001]{dok01}). In particular,
the variation with redshift of the composition of the galaxy
population in clusters suggests that S0 galaxies have come into being
fairly recently, say, since $z \sim 0.5$, apparently at the expense of
the spirals (\cite[Dressler et al. 1997; Fasano et
al. 2000]{dre97,fas00}).  The processes by which spirals transform
into S0 galaxies have been studied numerically (e.g. \cite[Moore et
al.  1998]{moo98}; \cite[Abadi et al. 1999]{aba99}). It appears that
impulsive encounters can indeed transform a spiral into an S0 galaxy,
by stripping a small fraction of the stellar disk, and by heating up
the disk.

We have used the data from the ENACS to investigate if the effects of
these processes are still detectable in present-day clusters. Our
analysis is based on an 'ensemble' cluster of $\sim 3000$ galaxies
built by combining 59 nearby ($0.035 \leq z \leq 0.098$) clusters from
the ENACS data-set (\cite[Katgert et al. 1996, 1998]{pk96,pk98}),
after suitable scaling of the galaxies projected clustercentric
distances and velocities (\cite[Biviano et al. 2002]{ab02}). For most
of these galaxies also spectral-type or morphological data are
available (\cite[Thomas 2002]{tt02}). We consider five populations of
cluster galaxies, selected on the basis of their different projected
phase-space distributions (\cite[Biviano et al. 2002]{ab02}). These
populations are: (i) galaxies in subclusters (herafter referred to as
the `Subs' class), and, among the galaxies outside subclusters: (ii)
the brightest ellipticals (with $M_R \leq -22+5 \log h$, the `Bright'
class), (iii) the other ellipticals together with the E/S0 and the S0
galaxies (the `Early' class), (iv) the early spirals (Sa--Sb; the
`S-early' class), and (v) the late spirals and irregulars (Sbc--Ir)
together with the emission-line galaxies (except those with early
morphology), hereafter globally referred to as the `S-late' class
galaxies.

The substructure distinction is based on a refined version of the
\cite{ds88} criterion. Galaxies in cold and/or moving substructures
were identified by the velocity dispersion and/or average velocity of
their surroundings, which are smaller than the global value, or
different from the cluster mean, respectively. With the significance
threshold that we used, about 25~\% of all galaxies are in
substructures.  The fraction of galaxies in substructures shows a
significant apparent decrease towards the cluster center. This is
probably a real effect, but there is likely to be a systematic
contribution from the selection algorithm (\cite[Biviano et
al. 2002]{ab02}).

\section{Clues from the spatial distributions}
\label{s-spatial}

The projected distributions of the 5 classes are shown in the
left-hand panel of Fig.~1, in the form of the smooth estimate of
$I(R)$ (using the {\em LOWESS} technique, see \cite[Gebhardt \& Fisher
1995; Katgert et al. 2004; Biviano \& Katgert
2004]{gf95;pk04;bk04}). Note that our cluster sample has a median
value $r_{200} \approx 1.2 \, h^{-1}$ Mpc.  The brightest ellipticals
(Bright-class galaxies) are most centrally concentrated. The
Early-class shows a flatter slope, however with a clear steepening
beyond $\approx 0.3 \, r_{200}$. The S-early- and S-late- classes
both avoid the central region within $\approx 0.1 \, r_{200}$, but in
addition the S-late-class shows an appreciable flattening of $I(R)$
below $\approx 0.3 \, r_{200}$. The Abel deprojection of $I(R)$ which
yields the 3-D number density profiles shows that both the S-early-
and S-late- classes are absent from the central region. Apparently,
galaxies of the S-early- and S-late- classes get destroyed or
transformed within $\approx 0.3 \, r_{200}$, as it is unlikely that
their phase-space distributions would prohibit them from entering the
central regions. Galaxies in substructures do not avoid the central
regions, but are less centrally concentrated than galaxies of the
Early-class.

\begin{figure}
\centering
\includegraphics{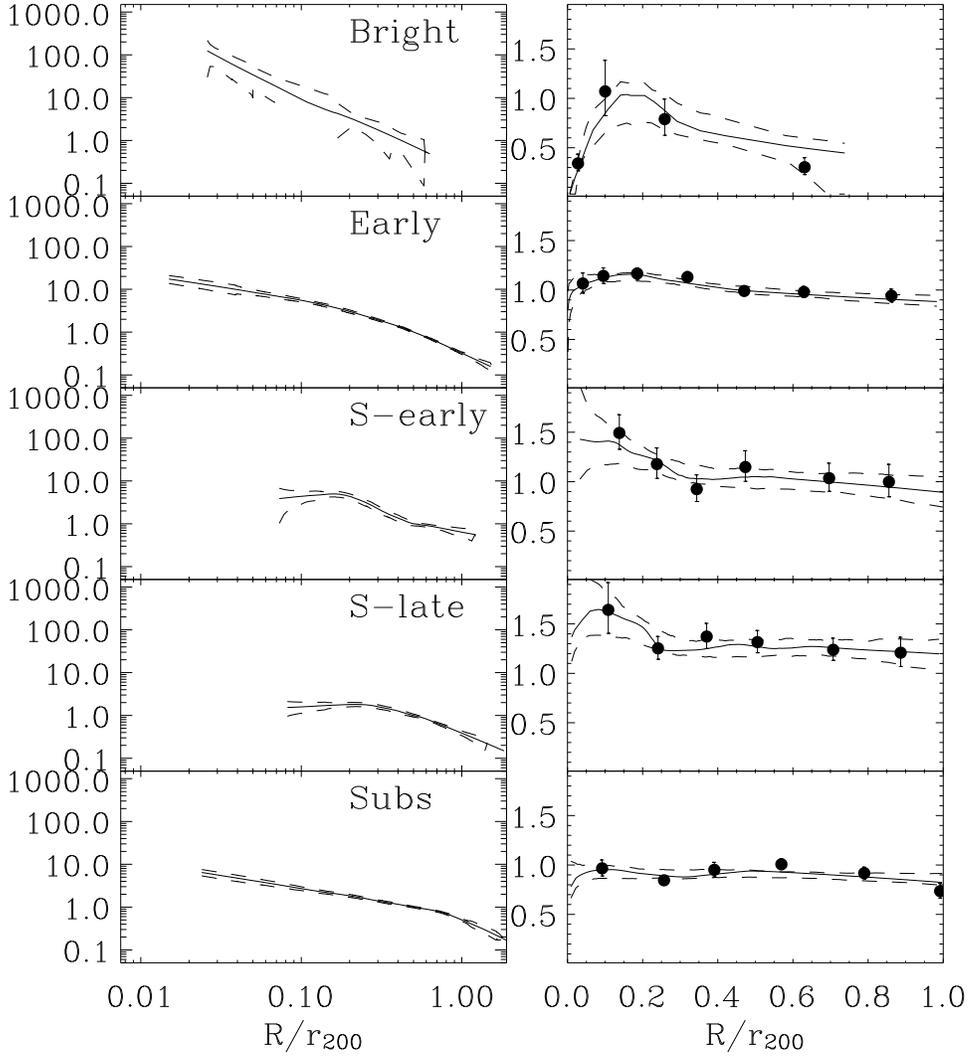}
\caption{The projected distributions (left) and the velocity
dispersion profiles (right) of the five galaxy classes. Dashed lines
indicate 1-$\sigma$ confidence levels. Dots (with 1-$\sigma$ error
bars) in the right-hand panel indicate binned estimates of the
velocity dispersion profiles. Units along the y-axis in the left-hand
panel are arbitrary. The y-axis in the right-hand panel is in units of
the cluster global line-of-sight velocity dispersion (on average,
$\approx 700$ km~s$^{-1}$ for the 59 clusters that compose the
ensemble cluster).}
\label{f-irvdp}
\end{figure}

Information on possible transformation relations between the galaxies
in the different classes may be gleaned from the morphology-radius
(MR) and, in particular, the morphology-density (MD) relation
(e.g. \cite[Dressler 1980]{dre80}). For the ENACS clusters those have
been discussed by \cite{tk04}, who considered ellipticals (including
the brightest) and S0 galaxies separately, in addition to the
S-early- and S-late-classes. The MR-relations show the dependence
on radius of the composition in terms of morphology, and are fully
consistent with the $I(R)$'s in the upper panel of Fig.~1. The
MD-relation gives the distribution of local projected density around
galaxies of the different classes. \cite{tk04} first used the original
definition of projected density, i.e. that based on the 10 nearest
neigbours, or $\Sigma_{10}$. However, $\Sigma_{10}$ contains a very
noticeable print-through of the average relation between projected
density and radius. As a result, the $\Sigma_{10}$ MD-relation is
highly correlated with the MR-relation (and thus, with the
$I(R)'s$). The ellipticals prefer the highest densities, and the
remaining three classes have indistinguishable $\Sigma_{10}$
distributions shifted towards lower densities (left-hand panel of
Fig.~2). From this perspective, the S0 galaxies and ellipticals behave
differently; part of this difference is due to the brightest
ellipticals being located at the clusters centres.

\cite{tk04} also used $\Sigma_{1}$, i.e. the projected density
based on the distance to the nearest neighbour. As expected,
$\Sigma_{1}$ is much noisier than $\Sigma_{10}$, but at the same time
it is much less correlated with projected radius, and therefore a less
biased measure of local projected density than $\Sigma_{10}$. That is
probably the reason why the $\Sigma_{1}$-distributions of S0 and
S-early galaxies are significantly different, with S0 galaxies
preferring higher densities than the S-early galaxies (right-hand
panel of Fig.~2).

\begin{figure}
\centering
\includegraphics{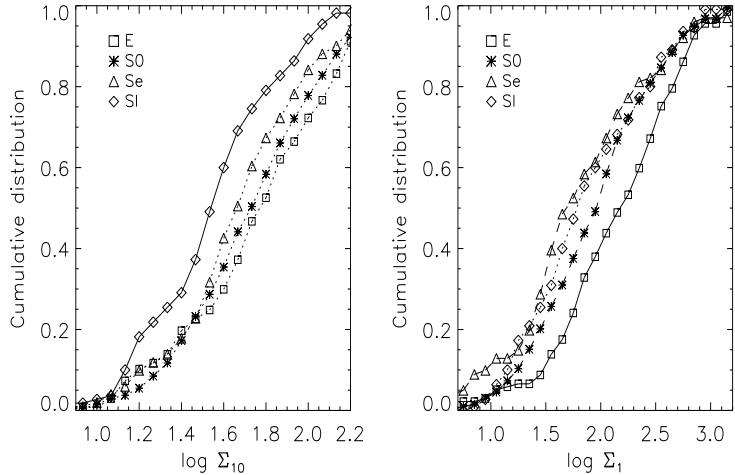}
\caption{The Morphology Density relations for $\Sigma_{10}$ (left)
and $\Sigma_{1}$ (right).}
\label{f-compos}
\end{figure}
\section{Clues from the kinematics}
\label{s-kinem}

\subsection{Isotropic orbit distributions?}

For a study of the kinematics of the various galaxy classes, we have
used the observed projected distributions $I(R)$ and velocity
dispersion profiles (vdp's, hereafter). The vdp's of the five
different classes are quite different (see right-hand panel of
Fig.~1). The Bright-class vdp is quite `cold', not only in the cluster
centre. The Early-class vdp is nearly flat. The vdp's of the S-early
and the S-late classes are rather similar, showing a steep decline
from the centre out to $\sim 0.3 \, r_{200}$, followed by a nearly
flat vdp. However, the S-late vdp is `hotter' than the S-early vdp at
all radii. Finally, the vdp of the Subs-class is cold and flat, even
flatter than the Early-class vdp.

In order to determine the kinematics of the different galaxy classes
we first need to determine the gravitational potential in which the
galaxies move. We have therefore first derived the mass-profile
$M(<r)$, using the data for the Early-class galaxies, assuming that
those have an isotropic distribution of orbits. That assumption is
justified by the shape of their line-of-sight velocity distribution,
from which \cite{pk04} concluded that, for the Early-class $0.8 \leq
\beta' \leq 1.05$, with
\begin{equation}
\beta' \equiv
(\overline{v_r^2}/\overline{v_t^2})^{1/2}, 
\end{equation}
where $\overline{v_r^2}$,
$\overline{v_t^2}$ are the mean squared components of the radial and
tangential velocity, respectively (i.e. isotropic orbits correspond to
$\beta'(r)=1$). Therefore the cluster mass profile, $M(<r)$, follows
from the {\em isotropic} Jeans equation (\cite[Binney \& Tremaine
1987]{bt87}) applied to the Early-class. The resulting $M(<r)$ is
rather similar to a NFW (\cite[Navarro et al. 1997]{nfw}) profile
with a concentration parameter $c=4.0_{-1.5}^{+2.7}$ (\cite[Katgert 
et al. 2004]{pk04}).

Next, this cluster $M(<r)$ was used to search for solutions of
dynamical equilibrium for the other four classes (see
\S~\ref{s-intro}). Given the cluster $M(<r)$ and the observed number
density profile of a given class of cluster galaxies, one can solve
the inverse Jeans equation (\cite[van der Marel 1994]{rv94}) for a
given $\beta'(r)$.  The resulting vdp is then compared, after the
usual Abel projection (\cite[Binney \& Tremaine 1987]{bt87}), with the
observed line-of-sight vdp, in a $\chi^2$ sense. If the two vdp's are
statistically indistinguishable, the $\beta'(r)$ profile is considered
acceptable. Instead of considering complicated $\beta'(r)$ profiles,
we tested whether the data were consistent with an isotropic orbit
distribution for the other galaxy classes, i.e. if $\beta'(r)=1$ is
acceptable. This appears to be the case for galaxies of the S-early
class, but not for those of the Bright, S-late, and Subs classes.

\subsection{The orbital anisotropy profiles}
\begin{figure}
\centering
\includegraphics{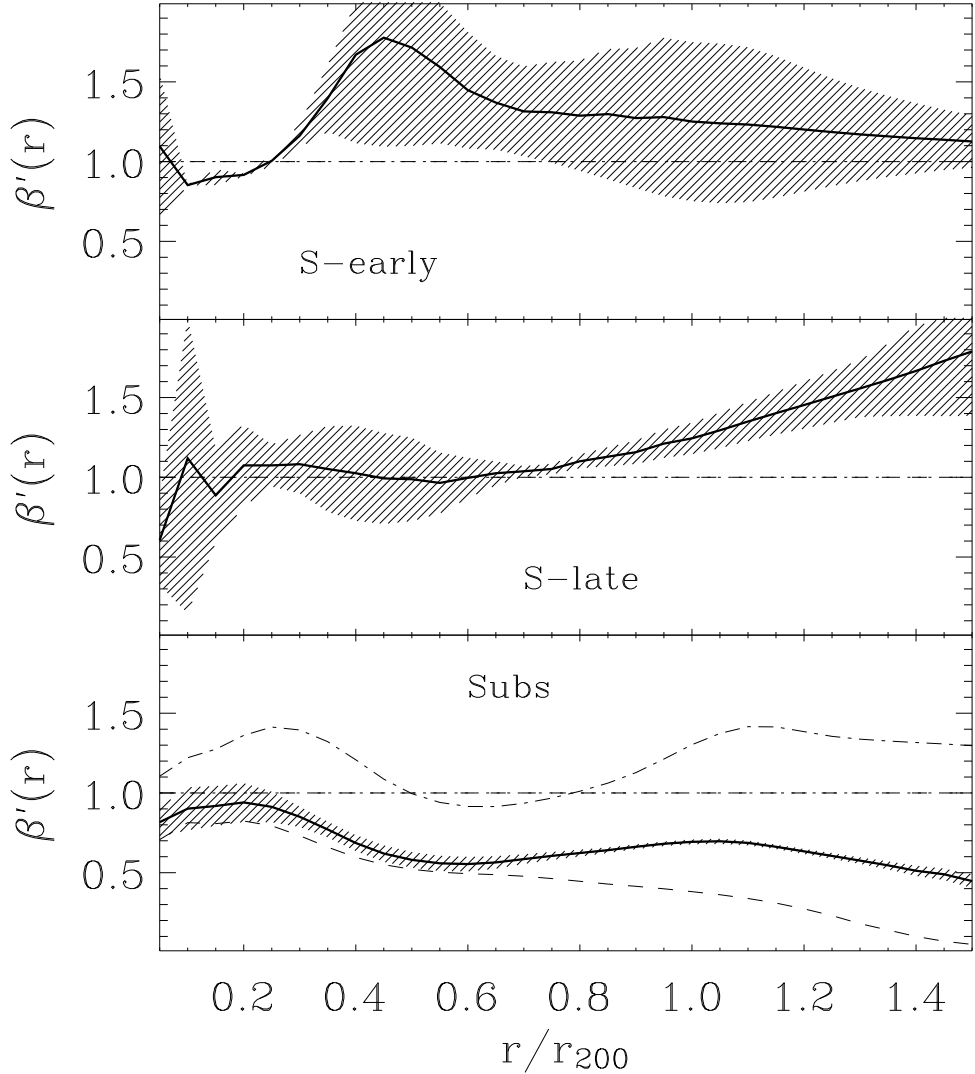}
\caption{Top panel: The velocity anisotropy profile, $\beta'(r) \equiv
(\overline{v_r^2}(r)/\overline{v_t^2}(r))^{1/2}$, of the S-early
class. The hatched region represents the approximate 1-$\sigma$
confidence interval. Mid panel: same as top panel, but for the S-late
class. Bottom panel: same as top panel, but for the Subs class. The
dashed line is the $\beta'$-profile obtained after subtracting in
quadrature the internal velocity dispersion of the subclusters from
the observed vdp. The dash-dotted line is the $\beta'$-profile
obtained after correcting the observed number density profile and vdp
of the Subs class for the systematic selection effects, according to a
plausible, but by no means unique, model.}
\label{f-beta}
\end{figure}

Since the assumption of (nearly) isotropic orbits is not valid for all
except the Early and, possibly, the S-early classes we have
investigated what the data tell us about the orbital anisotropy.
Instead of trying constant anisotropy (\cite[van der Marel et
al. 2000; \L okas \& Mamon 2003]{rv00,lm03}), or anisotropy models
(\cite[Carlberg et al. 1997]{rc97}), we determine $\beta'(r)$ directly
from the data, using the procedure that \cite{ss90} developed from
that of \cite{bm82}. The $\beta'(r)$ solutions are checked by
inverting the Jeans equation again (\cite[van der Marel 1994]{rv94}),
thereby deriving a vdp that can be compared with the observed one. The
derived vdp's always agree very well with the observed vdp's. We also
applied this check to the sample of Early-class galaxies, and indeed
recovered $\beta'(r)=1$ to within $\pm 15$\%.

Uncertainties on the $\beta'$-profiles were estimated by dividing each
galaxy sample into four subsamples, each containing a different half
of all the galaxies in the original sample. We then applied the 
procedure of \cite{ss90}
to each subsample. The rms of the four resulting
$\beta'$-profiles is an approximate, but probably conservative,
estimate of the uncertainty in the full sample $\beta'(r)$.

Physically acceptable solutions are found for the Early and three of
the four other populations of cluster galaxies. No acceptable solution
is found for the Bright class. This is most likely the consequence of
a failure of the collisionless Jeans equation in describing the
dynamics of these galaxies, which is probably affected by dissipative
processes, such as mergers and dynamical friction (see
\cite[Fusco-Femiano \& Menci 1998]{fm98} for a {\em collisional}
solution).

The $\beta'$-profile obtained for the S-early class increases rapidly
from $\beta' \simeq 1$ near the centre, to $\beta' \simeq 1.8$ at
$r/r_{200} \simeq 0.45$, and then smoothly decreases to $\simeq 1$
again at larger radii (see Fig.~3, top panel).
Taken at face value, this solution implies mild
radial anisotropy just outside the cluster centre. The partial radial
anisotropy at $r \simeq 0.45 \, r_{200}$ could result from a natural
selection effect. I.e., among S-early galaxies, those reaching the
high-density inner cluster region can avoid transforming into S0's
only if they move sufficiently fast. We should not overinterpret the
data, though, since the isotropic solution is acceptable for galaxies
of the S-early class, and the $\beta'$-profile is in fact consistent
with $\beta'(r)=1$ .

The $\beta'(r)$ profile of the S-late galaxies is not different from
unity out to $\simeq 0.7 \, r_{200}$, but then increases linearly with
radius reaching a radial anisotropy $\beta' \simeq 1.8$ at $r \simeq
1.5 \, r_{200}$ (see Fig.~3, mid panel). This $\beta'$-profile is
remarkably similar to those obtained for dark matter particles in
numerical simulations (\cite[Ghigna et al. 1998; Diaferio
1999]{sg98,ad99}). It is therefore tempting to conclude, by analogy,
that S-late galaxies, like dark matter particles in numerical
simulations, still retain memory of the process of (mostly radial)
gravitational infall along the filaments connecting to the cluster,
and are therefore recent arrivals into the cluster.

Finally, the data for the galaxies of the Subs class imply orbits with
substantial tangential anisotropy at all radii (see Fig.~3, solid line
in the bottom panel). However, two systematic effect must be accounted
for. The first one is the internal velocity dispersion of
subclusters. If we subtract an assumed internal velocity dispersion of
250~km~s$^{-1}$ from the observed vdp of Subs galaxies, the orbital
tangential anisotropy becomes even stronger (see Fig.~3, dashed line
in the bottom panel). The second effect is the radial variation of the
efficiency of the detection of Subs galaxies. Very simple modeling
shows that the efficiency of the selection method decreases towards
the cluster center. As a result, both the number density profile and
the vdp of the Subs class are affected. A simple first-order
correction of these profiles leads to a $\beta'$-profile which no
longer implies tangential orbits (see Fig.~3, dash-dotted line in the
bottom panel). A better understanding of the orbital characteristics
of subclusters thus requires either an improved, radially-unbiased,
selection method of galaxies in subclusters, or at least an improved
modeling of the biases.

\section{Summary}

The Early-class galaxies (ellipticals and S0 galaxies) in clusters
have a distribution of line-of-sight velocities that is consistent
with an isotropic orbit distribution. These galaxies were therefore
used to derive the mass profile of the ENACS ensemble cluster. Using
this mass profile, we constrained the orbits of cluster galaxies of
other classes. From the distribution and velocity dispersion profile
of the S-early-class we conclude that these may also have isotropic
orbits, although they show an apparent radial anisotropy around
$\approx 0.45 \, r_{200}$. The latter may be a result of the process
by which the majority of the S-early-galaxies are thought to have
transformed into S0 galaxies through impulsive encounters. The fact
that the local density around S-early-galaxies is smaller than that
around S0 galaxies provides further support for this
interpretation. The S-late-class has mildly radial orbits outside
$\approx 0.7 \, r_{200}$, and are probably still falling in. Their
absence from the central region most likely indicates that they are
tidally disrupted on their crossing of the cluster core. Finally, the
galaxies in substructures apparently are on tangential orbits,
although isotropic (or mildly radial) orbits cannot be excluded at
present.

\begin{acknowledgments}
We wish to thank the Local Organizing Committee for a very
enjoyable and interesting meeting. We acknowledge useful discussions
with Alain Mazure and Tom Thomas.
\end{acknowledgments}


\begin{thebibliography}{}

\bibitem[Abadi et al. (1999)]{aba99} 
Abadi, M. G., Moore, B., \& Bower, R., G. 1999, {\em MNRAS,} {\bf 308,} 947.

\bibitem[Binney \& Mamon (1982)]{bm82} 
Binney, J., \& Mamon, G. 1982, {\em MNRAS,} {\bf 200,} 361.

\bibitem[Binney \& Tremaine (1987)]{bt87}
Binney, J. \& Tremaine, S. 1987, {\em ``Galactic Dynamics''}
(Princeton Univ. Press).

\bibitem[Biviano et al. (2002)]{ab02}
Biviano, A., Katgert, P., Thomas, T., \& Adami, C. 2002, {\em A\&A,} 
{\bf 387,} 8.

\bibitem[Biviano \& Katgert (2004)]{bk04}
Biviano, A., \& Katgert, P. 2004, {\em in preparation.}

\bibitem[Carlberg et al. (1997)]{rc97}
Carlberg, R.G., Yee, H.K.C., \& Ellingson, E. 1997, {\em ApJ,} {\bf 478,} 462.

\bibitem[Diaferio (1999)]{ad99}
Diaferio, A. 1999, {\em MNRAS,} {\bf 309,} 610.

\bibitem[Dressler (1980)]{dre80} 
Dressler, A. 1980, {\em ApJ,} {\bf 236,} 351.

\bibitem[Dressler \& Shectman (1988)]{ds88}
Dressler, A. \& Shectman, S.A. 1988, {\em AJ,} {\bf 95,} 985.

\bibitem[Dressler et al. (1997)]{dre97} 
Dressler, A., Oemler, A. Jr., Couch, W. J., et al. 1997, 
{\em ApJ,} {\bf 490,} 577.

\bibitem[Fasano et al. (2000)]{fas00} 
Fasano, G., Poggianti, B. M., Couch, W. J., et al. 2000, {\em ApJ,} 
{\bf 542,} 673.

\bibitem[Fusco-Femiano \& Menci (1998)]{fm98}
Fusco-Femiano, R. \& Menci, N. 1998, {\em ApJ,} {\bf 498,} 95.

\bibitem[Gebhardt \& Fisher (1995)]{gf95}
Gebhardt, K. \& Fischer, P. 1995, {\em AJ,} {\bf 109,} 209

\bibitem[Ghigna et al. (1998)]{sg98}
Ghigna, S., Moore, B., Governato, F., et al. 1998, {\em MNRAS,} {\bf 300,} 146.

\bibitem[Katgert et al. (1996)]{pk96}
Katgert, P., Mazure, A., Perea, J., et al. 1996, {\em A\&A,} {\bf 310,} 8.

\bibitem[Katgert et al. (1998)]{pk98}
Katgert, P., Mazure, A., den Hartog, R., et al. 1998, {\em A\&AS,} {\bf 129,} 399.

\bibitem[Katgert et al. (2004)]{pk04}
Katgert, P., Biviano, A., \& Mazure, A. 2004, {\em ApJ,} {\bf 600,} 657.

\bibitem[\L okas \& Mamon (2003)]{lm03}
\L okas, E. \& Mamon, G. 2003, {\em MNRAS,} {\bf 343,} 401.

\bibitem[Moore et al. (1998)]{moo98} 
Moore, B., Lake, G., \& Katz, N. 1998, {\em ApJ,} {\bf 495,} 139.

\bibitem[Navarro et al. (1997)]{nfw}
Navarro, J.F., Frenk, C.S. \& White, S.D.M. 1997, {\em ApJ,} {\bf 490,} 493.

\bibitem[Solanes \& Salvador-Sol\'e (1990)]{ss90}
Solanes, J.M., \& Salvador-Sol\'e 1990, {\em A\&A,} {\bf 234,} 93.

\bibitem[Thomas (2002)]{tt02}
Thomas, T. 2002, {\em PhD thesis, Leiden Observatory.}

\bibitem[Thomas \& Katgert (2004)]{tk04}
Thomas, T., \& Katgert, P. 2004, {\em in preparation.}

\bibitem[van der Marel (1994)]{rv94}
van der Marel, R. 1994, {\em MNRAS,} {\bf 270,} 271.

\bibitem[van der Marel et al. (2000)]{rv00}
van der Marel, R., Magorrian, J., Carlberg, R., et al. 2000, {\em AJ,} {\bf 119,} 2038.

\bibitem[van Dokkum and Franx (2001)]{dok01} 
van Dokkum, P. G., \& Franx, M. 2001, {\em ApJ,} {\bf 553,} 90.

\end{thebibliography}
\end{document}